\documentclass[prb,twocolumn,showpacs,aps,floats,floatfix,superscriptaddress]{revtex4}
\usepackage{epsfig}

\begin{document}
\title{Role of electrostatic interactions in the assembly of empty spherical viral capsids}
\author{Antonio \v{S}iber}
\email{asiber@ifs.hr}
\affiliation{Department of Theoretical Physics, Jo\v{z}ef Stefan Institute, SI-1000 Ljubljana, Slovenia}
\affiliation{Institute of Physics, P.O. Box 304, 10001 Zagreb, Croatia}

\author{Rudolf Podgornik}
\email{rudolf.podgornik@fmf.uni-lj.si}
\affiliation{Department of Theoretical Physics, Jo\v{z}ef Stefan Institute, SI-1000 Ljubljana, Slovenia}
\affiliation{Department of Physics, University of Ljubljana, SI-1000 Ljubljana, Slovenia}

\begin{abstract}
We examine the role of electrostatic interactions in the assembly of empty spherical viral capsids. 
The charges on the protein subunits that make the viral capsid mutually interact and are expected to 
yield electrostatic repulsion acting against the assembly of capsids. Thus, attractive protein-protein 
interactions of non-electrostatic origin must act to enable the capsid formation. We investigate 
whether the interplay of repulsive electrostatic and attractive interactions between the protein subunits 
can result in the formation of spherical viral capsids of a preferred radius. For this to be the case, 
we find that the attractive interactions must depend on the angle between 
the neighboring protein subunits (i.e. on the mean curvature of the viral capsid) so that a particular 
angle(s) is (are) preferred energywise. Our results for the electrostatic contributions to 
energetics of viral capsids nicely correlate with recent experimental 
determinations of the energetics of protein-protein contacts in Hepatitis B virus [P. Ceres and 
A. Zlotnick, Biochemistry {\bf 41}, 11525 (2002).].
\end{abstract}
\date{\today}
\maketitle

\section{Introduction}

Many types of viruses can spontaneously assemble from the proteins that 
make the viral coating (capsid) and the viral genetic material. This was first 
demonstrated in the work of Fraenkel-Conrat and Williams \cite{Fraenkel} where they showed that 
fully infectious tobacco mosaic viruses can be spontaneously 
reconstituted from the individual proteins that make the viral coating and the viral 
RNA molecules.

It may be argued that the precision of the viral assembly is guided by an interplay 
between the properties of the RNA/DNA (bio-polyelectrolyte) and the capsid proteins, such as the 
total bio-polyelectrolyte length, the effective charge on the proteins and on the bio-polyelectrolyte, and possibly some specific interaction 
acting between the RNA/DNA and the proteins. However, in some types of 
viruses, the genetic material is not necessary for the capsid assembly and {\em empty} 
viral capsids can be assembled in the absence of viral bio-polyelectrolyte 
\cite{Iwasaki,zlotnickhepb,zlotnickCCMV,Casparpolio,Ganser}, at least when the amount of salt in the 
solution is large enough \cite{Ganser}. It is rather 
intriguing that the thus assembled capsids are often highly monodisperse \cite{Iwasaki,zlotnickhepb} 
which suggests that there is some regulating factor, independent of the viral RNA/DNA, that 
favors capsids of particular size. This is the effect that we investigate in this article and 
the motivation for our study.

In principle, viral proteins can assemble in a variety of capsids with different shape and size 
\cite{Casparpolio,kellenberger,bruinthermo}. We shall concentrate on the nearly spherical viral capsids whose structure can be described 
within the so-called Caspar-Klug quasiequivalent construction \cite{casparklug}. 
In more mathematical terms, 
the viral capsids that we consider are icosadeltahedral, i.e. they can be 
mapped onto triangulations of the sphere with the icosahedral ''backbone'' 
(see e.g. Refs. \onlinecite{kellenberger,zlotstruct,Siber1} for more details). The different triangulations can 
be described using the notion of the so-called ${\cal T}$-number. The number of protein 
subunits ($N$, which may also be protein dimers, such is e.g. the case in Hepatitis B virus) 
in a quasiequivalent viral capsid is 
\begin{equation}
N=60 {\cal T}, 
\label{eq:protnumber}
\end{equation}
where ${\cal T}=1, 3, 4, 7, 9, ...$, i.e. ${\cal T}=h^2 + hk + k^2$ and $h$ and $k$ are non-negative integers \cite{Siber1}.

In view of the potential polymorphism of the viral protein assemblies which is also 
observed experimentally, especially in some viruses \cite{Ganser,Auravirus,P22}, it is rather 
surprising that empty capsids of most simple viruses precisely assemble in capsids 
whose $T$-number is the same as in fully functional capsids that contain the 
viral RNA/DNA \cite{Iwasaki,zlotnickhepb}. Even when 
several differently sized empty capsids do form \cite{zlotnickCCMV,Casparpolio,P22}, 
these represent only a tiny subset of all of the 
imaginable capsid formations. Since the viral proteins have 
certain charge in the solution \cite{Kegel,TMVKegel}, it is tempting to assume that the repulsive 
electrostatic interactions compete with some other attractive interactions between 
the viral proteins in such a way that the total free energy of the capsid is 
minimal exactly for the capsid of the observed radius. The attractive interactions 
between the viral proteins could be of different origins (e.g. van der Waals interactions, hydrophobic interactions 
or chemical bonding). However, it has been experimentally 
demonstrated that the binding energy of the two viral proteins increases with 
temperature \cite{zlotnickhepb}, which strongly suggests that 
the attractive protein interactions are dominated by either the zero-frequency term of van der Waals 
interactions \cite{parsegian} and/or by hydrophobic interactions \cite{Kegel}. They should thus 
be rather local and proportional to the area ''buried'' in the protein-protein contacts.

The competition between hydrophobic and electrostatic interactions in viral capsids 
has been theoretically investigated before \cite{Kegel,bruin1,zandi1}. The emphasis there, however, 
was mostly on kinetics of viral assembly, i.e. on the classical nucleation 
theory and the mass action law. Furthermore, the 
electrostatic interactions were modeled as the asymptotic form of the Debye-H\"{u}ckel 
(DH) approximation, in the regime where $R / \lambda_{DH} \gg 1$, where $R$ is the capsid 
radius and $\lambda_{DH}$ the DH screening length that is proportional to inverse square 
root of the salt concentration. It is the aim of this article 
to reexamine in more detail the role of electrostatic interactions in the capsid 
assembly, to investigate whether they can act to regulate the capsid size (radius) and 
whether they can prevent the capsid formation. To this end, we shall introduce a model 
of a viral capsid that is more realistic and more complex than the ones used in 
previous studies. We shall go beyond the DH approximation and consider 
all regimes in $R / \lambda_{DH}$. 
We shall also estimate the contribution of the electrostatic interactions to the total 
binding energy of the capsid. 

There are two different models of capsid electrostatics that 
we consider. Section \ref{sec:model1} describes the simpler of these models that 
treats the capsid as uniformly charged, infinitely thin sphere whose charge density is fixed by the 
total charge on the capsid protein in the solution. This, rather simplified, model of capsid 
allows us to relate our results with those previously published \cite{Kegel}. The aim of 
Section \ref{sec:model1} is twofold. First, we shall clearly demonstrate that the previously 
derived expressions for the capsid energetics \cite{Kegel} have limited validity, even in 
physiological conditions. The theory and results presented in Section \ref{sec:model1} also constitute a 
good prelude for a more elaborate 
electrostatic model of a viral capsid presented in Section \ref{sec:model2}. 
There, the capsid is modeled as a dielectric medium contained in between the two infinitely 
thin spheres, each of which has a certain prescribed charge density. In Section 
\ref{sec:application} we attempt to relate our theoretical predictions to 
experimental results on the viral energetics \cite{zlotnickhepb}. Section 
\ref{sec:seclimit} discusses limitations of our models of viral capsids.

\section{Viral capsid as an infinitely thin, uniformly charged spherical shell: model I}
\label{sec:model1}

\subsection{Free energy of empty capsids in salty solution}

We first approximate the icosadeltahedral capsid as the perfect sphere of radius $R$ whose charge is uniformly 
distributed on the surface, so that the surface charge density is $\sigma$. The theory of 
elasticity applied to icosadeltahedral shells predicts that the viral capsids are aspherical, 
the more so the larger their mean radius (this is also in agreement 
with experimental data) \cite{Siber1,Nelson1}. The asphericity arises from the so-called ''buckling'' of the 
capsid around the pentameric protein aggregates \cite{Nelson1,Siber1}. The experimental 
data on virus shapes \cite{Baker} suggests that the virus surface may in fact be very corrugated, but 
it is difficult to assess the corrugation and asphericity of the corresponding protein charge distribution. 
For our purposes we neglect the capsid asphericity which is, on the basis of continuum theory of Lidmar 
{\sl et al} \cite{Nelson1}, expected to be small especially in viruses of small radii (the validity of 
the theory may, however, be questionable for small viruses; see Sec. \ref{sec:seclimit}). 
Under this approximation, the charges on the capsid, 
together with the (monovalent) salt ions in the solution whose bulk concentration is $c_0$, 
give rise to spherically symmetric electrostatic potential $\Phi(r)$, so that the problem is effectively one-dimensional. 

The free energy of the system (proteins and salt ions in the solution) in the mean-field (Poisson-Boltzmann) approximation can be expressed 
as a functional of the electrostatic potential as \cite{Andelman1}
\begin{equation}
F = \int d^3 r \left [ f_{el}(r) + f_{ions}(r) \right ] + F_{boundary},
\label{eq:fefunc}
\end{equation}
where
\begin{equation}
f_{el}(r) = e c^{+} \Phi - e c^{-} \Phi - \frac{\epsilon_0 \epsilon_r}{2} |\nabla \Phi|^2,
\label{eq:electrostatf}
\end{equation} 
and
\begin{eqnarray}
f_{ions}(r) &=& \sum_{i= \pm} \left \{ 
\frac{1}{\beta} \left [ 
c^{i}(r) \ln c^{i}(r) - c^{i}(r) \right. \right . \nonumber \\
&-& \left . \left . \left( 
c_0^{i} \ln c_{0}^{i} - c_{0}^{i}
\right)
\right ]
- \mu^{i} \left[ 
c^{i}(r) - c_{0}^{i}
\right]
\right \},
\label{eq:fions}
\end{eqnarray}
where $F_{boundary}$ is the boundary contributions arising from the discontinuity of the potential at the 
capsid [see Eqs. (\ref{eq:boundsigma}) and (\ref{eq:totalfree}) below]. The free energy is the sum of the 
electrostatic energy  of charge in the potential ($f_{el}$), and the salt ions configurational/entropy 
contribution ($f_{ions}$). This {\sl ansatz} is correct as long as the counterions are of low valency 
and/or the charge density on the capsid is not too large, leading to the {\sl s.c.} weak coupling-regime 
which is properly captured by the Poisson-Boltzmann theory \cite{Naji}.
 
In the equations above, $e$ is the electron charge, 
$\epsilon_r$ is the relative permittivity 
of the solvent (water, $\epsilon_r=80$), $\epsilon_0$ is the vacuum permittivity, 
$\beta = (k_B T)^{-1}$, $T$ is the temperature ($T$=300 K), and $k_B$ the Boltzmann constant, 
$c^{+}(r)$ and $c^{-}(r)$ are the concentrations of $+$ and $-$ ions in the solution 
and $\mu^{+}$ and $\mu^{-}$ are their respective chemical potentials. 
We shall consider dilute capsid protein solutions and shall thus 
neglect the influence of counterions on the potential. This also 
means that the bulk concentrations of positive and negative ions are equal, 
$c_{0}^{+} = c_{0}^{-} = c_0$.

The variation of the functional Eq. (\ref{eq:fefunc}) with 
respect to fields $\Phi$ and $c^{i}$ yields the following equations:
\begin{equation}
c^{\pm} (r) = c_0 \exp [\mp e \beta \Phi (r) ],
\end{equation}
and
\begin{equation}
\epsilon_0 \epsilon_r \nabla ^2 \Phi(r) = 2 e c_0 \sinh [\beta e \Phi(r) ].
\label{eq:PBE1}
\end{equation}
Equation (\ref{eq:PBE1}) is the Poisson-Bolzmann equation for the potential field. 
Its mean field character is a consequence of approximating the salt ions as 
the ideal gas [Eq. (\ref{eq:fions})]. One additionally needs to specify the boundary condition of the  
$\Phi$ field.  For the assembly of virus-like particles, the prescribed surface charge density ($\sigma$) 
is the most realistic 
boundary condition and regardless of the capsid size, i.e. the number of protein 
subunits it contains, the surface charge density of capsid should be the same. We 
shall also assume that the charge on the particular protein does not depend 
on the salt concentration (but see Ref. \onlinecite{TMVKegel} for the dependence 
of protein charge on the pH value of solution) and will thus exclude charge regulation boundary 
condition from our considerations \cite{ninham}. Requiring that the surface charge density of 
the bounding sphere is fixed, yields
\begin{equation}
\left. \frac{\partial \Phi(r)}{\partial r} \right |_{r = R^-} - 
\left. \frac{\partial \Phi(r)}{\partial r} \right |_{r = R^+} = \frac{\sigma}{\epsilon_0 \epsilon_r}.
\label{eq:boundsigma}
\end{equation}
The derivative of the potential (or electric field) displays a discontinuity 
at the bounding sphere (see Fig. \ref{fig:figempty1}). This is of course 
a consequence of the fact that there is 
a charge ($Q=4 \pi \sigma R^2$) on the sphere. The electrostatic energy of this charge is equal to 
$F_{boundary} = Q \Phi(R)$ and the total electrostatic free energy of the system is thus 
\begin{equation}
F = \int [ f_{el}(r) + f_{ions}(r)] d^3 r + Q \Phi(R).
\label{eq:totalfree}
\end{equation}
Note that in the limit $c_0 \rightarrow 0$ (the Coulomb limit), the functional reduces to 
\begin{equation}
F =  - \frac{\epsilon_0 \epsilon_r}{2} \int |\nabla \Phi|^2  d^3 r + 
Q \Phi(R) = \frac{Q \Phi(R)}{2},
\end{equation}
which is simply the self-energy of the (completely unscreened) charge on the sphere.

\subsection{Numerical solutions of the model and analytic approximations in the Debye-H\"{u}ckel limit}

Equation (\ref{eq:PBE1}) is a nonlinear differential equation which we solve numerically. We first 
discretize the radial coordinate both within the capsid and outside it, so that the 
total number of points is typically around 600 \cite{primjedba}. The intervals in 
radial coordinates are not the same in the two regions, and have to be chosen so that the potential 
at the point with largest radial coordinate (outside the sphere) is very nearly close to zero. 
At each of the points we consider the value of the electrostatic potential. The 
sum of square deviation of the potential from the values required by the differential equation (\ref{eq:PBE1}) 
at the discrete set of radial coordinates is minimized \cite{minpack} until a desired numerical 
accuracy is achieved. Similar numerical procedure has also been used in Ref. \onlinecite{Andelman1}.

\begin{figure}[ht]
\centerline{
\epsfig {file=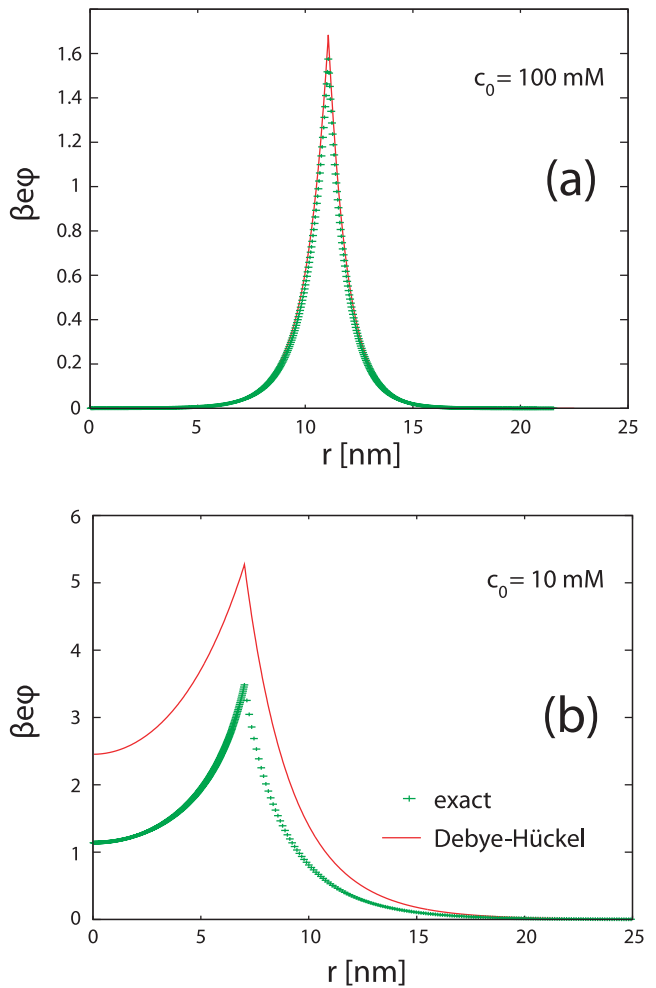,width=8.5cm}
}
\caption{Numerically exact solution for the potential $\Phi(r)$ (symbols) and the DH solution (lines). 
The (mono-valent) salt concentrations are 100 mM and 10 mM and the capsid radii are $R$=11.07 nm and 
$R$=7.02 nm in panels (a) and (b), respectively. The surface density of capsid charge is 
$\sigma$ = 0.4 $e$/nm$^2$.}
\label{fig:figempty1}
\end{figure}

Additional insight can be obtained in the regime when $e \beta |\Phi| \ll 1$. In that 
case (the DH approximation) the equation can be linearized and analytically solved. 
The solution within the capsid is
\begin{equation}
\Phi(r) = 
\frac{Q \sinh (\kappa_{DH} r)}{4 \pi r \epsilon_0 \epsilon_r \kappa_{DH} R \left[ \sinh (\kappa_{DH} R) + \cosh (\kappa_{DH} R)\right]},
\label{eq:DHpotential0} 
\end{equation}
and outside the capsid
\begin{equation}
\Phi(r) = 
\frac{Q \exp [-\kappa_{DH} (r-R)] }{4 \pi r \epsilon_0 \epsilon_r \kappa_{DH} R \left[ 1 + \coth (\kappa_{DH} R)\right]}, 
\label{eq:DHpotential}
\end{equation}
where $\kappa_{DH} = 1 / \lambda_{DH} = \sqrt{(2 e^2 c_0) / (\epsilon_0 \epsilon_r k_B T) }$ is the inverse DH screening length.  
Comparison of the numerical and the DH solution from the above two equations is shown in Fig. \ref{fig:figempty1}.
%%%%%%%%%%%%%%%%%%%%%
%%%%%%%%%%%%%%%%%%%%%%
Note how the DH approximation fails when the potential at the capsid acquires high values, as is the 
case in low-salt solutions. The surface charge density chosen in the displayed result ($\sigma$ = 0.4 $e$/nm$^2$) 
should be realistic for the assembly of virus-like particles at neutral pH \cite{Kegel,bruin1,TMVKegel}.

The free energy of the assembled capsid in salty solution can be calculated from Eq. (\ref{eq:totalfree}) once the 
electrostatic potential has been obtained. It is instructive to examine the Coulomb case first. In this 
situation, the total free energy is simply the electrostatic self-energy of the capsid, 
\begin{equation}
F_{Coulomb} = \frac{Q \Phi(R)}{2} = \frac{2 R^3 \pi \sigma^2}{\epsilon_0 \epsilon_r}.
\label{eq:nosaltfree}
\end{equation}
For a given surface charge density, the free energy should scale with the third power of the capsid radius. 
If one assumes that the attractive protein-protein interaction (whatever its source may be) is local, 
one can write
\begin{equation}
F_{adhesion} = -\frac{m \gamma}{2} N,
\label{eq:adhesimp}
\end{equation}
where $N$ is given by Eq.(\ref{eq:protnumber}), $m$ is the mean number of contacts that a 
particular protein makes with its neighbors ($m \sim 3$), and $\gamma$ is the negative 
free energy of the protein-protein contact 
($\gamma > 0$). Thus, the contribution of $F_{adhesion}$ to the 
total capsid energy should scale as $-R^2$, since the attractive interaction is 
simply proportional to the number of protein subunits. The combination of the unscreened electrostatic self-repulsion 
and the attractive adhesion interaction should thus have a minimum at some radius. This observation could apparently 
explain the origin of the monodispersity of capsid radii without the need to introduce e.g. the 
spontaneous curvature contribution to the total energy of the capsid. However, as shall be shown 
below, these considerations profoundly change when one considers the role of salt in the capsid assembly.

\begin{figure}[ht]
\centerline{
\epsfig {file=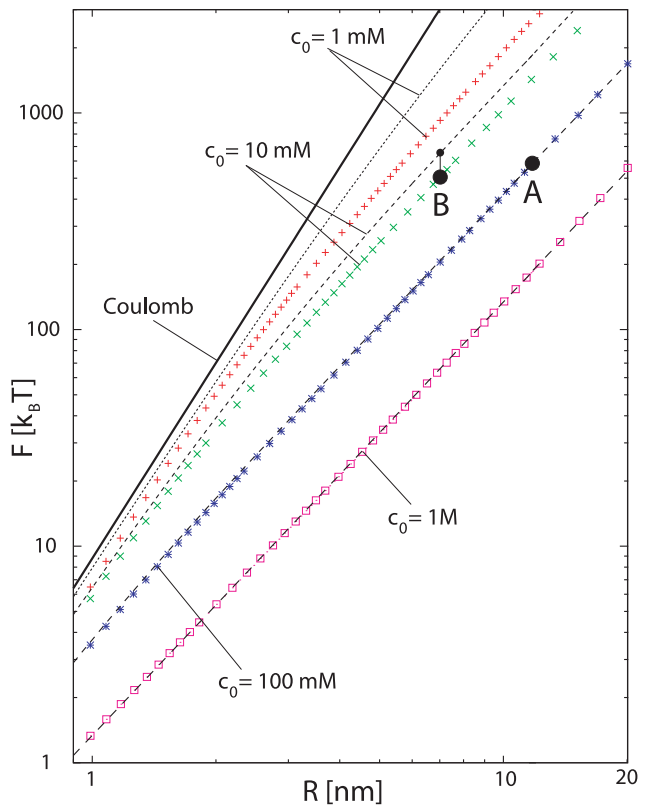,width=8.4cm}
}
\caption{Free energies of the system (capsid + salt) as a function of the capsid radius for several different salt 
concentrations as indicated in the body of the figure. The capsid surface charge density is 
$\sigma$ = 0.4 $e$/nm$^2$. The symbols denote the numerically obtained (exact) results, while 
the dashed lines are the DH approximations to the free energy (the length of dashes increases with the 
salt concentration). The full line denotes the 
free energy in the Coulomb regime, Eq. (\ref{eq:nosaltfree}). The points denoted by A and 
B correspond to potential profiles that are shown in panels (a) and (b) of Fig. \ref{fig:figempty1}, 
respectively.}
\label{fig:figempty2}
\end{figure}

\begin{figure}[ht]
\centerline{
\epsfig {file=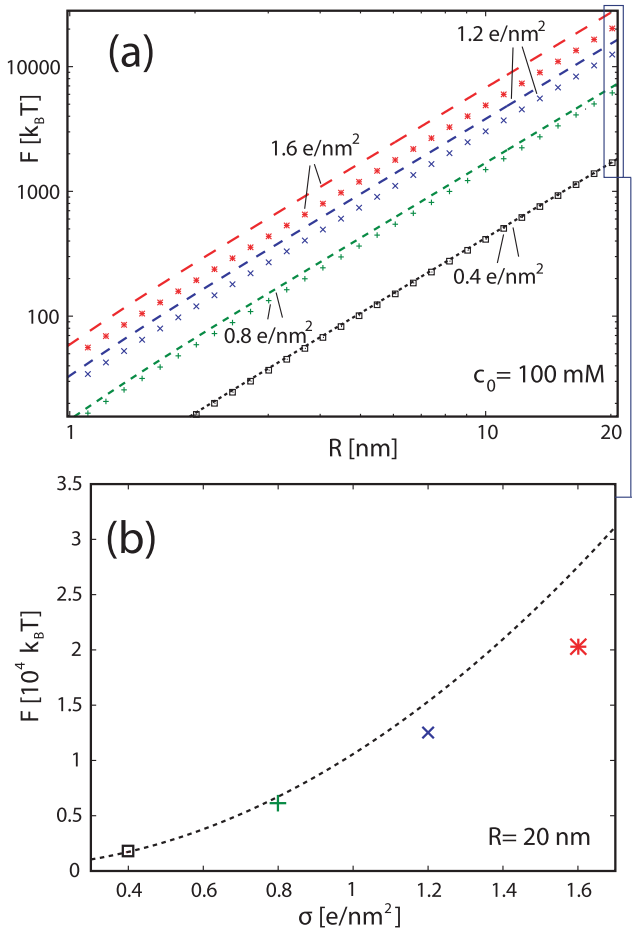,width=8.4cm}
}
\caption{Panel (a): Free energies of the system (capsid + salt) for $c_0$=100 mM as a function of the capsid radius for several different 
capsid charge densities as indicated in the body of the figure. The symbols denote the numerically obtained (exact) results, while 
the dashed lines are the DH approximations to the free energy (the length of dashes increases with the 
capsid charge density). Panel (b): Comparison of numerically exact results (symbols) with the DH approximation (dashed line) 
for capsid radius $R=$20 nm.}
\label{fig:figempty3}
\end{figure}

The free energies of the capsids in the solutions with several different mono-valent salt concentrations 
are shown in Fig. \ref{fig:figempty2}.
%%%%%%%%%%%%%%%%%%%%%
%%%%%%%%%%%%%%%%%%%%%%
It is obvious that even for quite low salt concentrations ($c_0 =$1 mM) the Coulomb prediction is 
not valid. Furthermore, even the functional dependence of the free energy on the capsid radius is 
different from the prediction of Eq. (\ref{eq:nosaltfree}). In the DH approximation, the 
free energy of the system can be shown to be given by 
\begin{equation}
F_{DH} = \frac{Q \Phi(R)}{2},
\end{equation}
somewhat deceivingly looking like the expression in Eq. (\ref{eq:nosaltfree}) since 
the potential at the capsid, $\Phi(R)$ is to be obtained as the solution of the linearized Poisson-Boltzmann 
equation, {\sl i.e.} it is profoundly influenced by salt ions. Using Eq. (\ref{eq:DHpotential0}), one obtains that
\begin{equation}
F_{DH} = \frac{2 \pi \sigma^2 R^2 }{\kappa_{DH} \epsilon_0 \epsilon_r [1 + \coth(\kappa_{DH} R)] }.
\label{eq:freeDHtot}
\end{equation}
When $\kappa_{DH} R \gg 1$, the equation reduces to
\begin{equation}
\lim _{\kappa_{DH} R \gg 1} F_{DH} = \frac{\pi \sigma^2 R^2 }{\kappa_{DH} \epsilon_0 \epsilon_r},
\label{eq:limitlarge}
\end{equation}
which shows that for $\kappa_{DH} R \gg 1$ the free energy scales with the {\em second} power 
of the capsid radius in clear contrast with the prediction obtained in the Coulomb regime, 
Eq. (\ref{eq:nosaltfree}). Exactly the same relation was obtained in Ref. \onlinecite{Kegel} 
[see their Eq. (4)] by using a different approach, but note that its regime of validity is limited by the fact that {\em (i)} 
it was derived by linearizing the Poisson-Boltzmann equation, and that {\em (ii)} 
it holds only when $\kappa_{DH} R \gg 1$. Thus, for very low salt concentrations, 
the validity of Eq. (\ref{eq:limitlarge}) is severely limited, but even for 
moderate salt concentration of $c_0$ = 10 mM (with 
$\sigma$ chosen as before), the DH approximation overestimates the free energies 
at $R \sim$ 10 nm by about 50 \%. For 
larger surface densities, the DH approximation is worse, and for sufficiently large capsid charge density 
it becomes erroneous even in the physiological salt regime ($c_0 \sim$ 100 mM). This is illustrated in Fig. \ref{fig:figempty3} 
which displays the free energies of the assembled capsids at $c_0$ = 100 mM for several different 
values of the effective capsid charge density ($\sigma = 0.4, 0.8, 1.2, 1.6$ $e$/nm$^2$).
%%%%%%%%%%%%%%%%%%%%%
%%%%%%%%%%%%%%%%%%%%%%
 
Interestingly enough, in the opposite regime of capsid sizes, i.e. 
when $\kappa_{DH} R \ll 1$, the DH expression for the system free energy reduces to
\begin{equation}
\lim _{\kappa_{DH} R \ll 1} F_{DH} = \frac{2 R^3 \pi \sigma^2}{\epsilon_0 \epsilon_r (1 + \kappa_{DH} R )},
\end{equation}
which clearly tends toward the (unscreened) Coulomb expression for the free energy when $\kappa_{DH} R \rightarrow 0$. 
It should be noted here that there is in fact an upper limit on $\lambda_{DH}$ set by the concentration 
of counterions that the proteins release in the solution, so that the limit $\lambda_{DH} \rightarrow \infty$ 
(or $\kappa_{DH} \rightarrow 0$) should be considered with caution.

The numerically exact solutions also display the scaling laws predicted by the DH approximation, 
i.e. the free energy scales as $R^3$ for $\kappa_{DH} R \ll 1$ and as $R^2$ for $\kappa_{DH} R \gg 1$ (see 
particularly the $c_0$=1 mM case in Fig. \ref{fig:figempty2}). One important difference though, is 
an onset of $R^2$ scaling behavior for smaller values of $R$ than predicted by the DH calculation, 
especially when the salt concentration is low. The same scaling behaviors have been obtained in a DH 
approximation for a more complicated model of a spherical viral capsid whose charge 
distribution is nonuniform and carries a signature of its icosahedral symmetry \cite{Marzec}.

\section{Viral capsid as a dielectric medium contained in between the two uniformly 
charged infinitely thin spherical shells: model II}
\label{sec:model2}

\begin{figure}[ht]
\centerline{
\epsfig {file=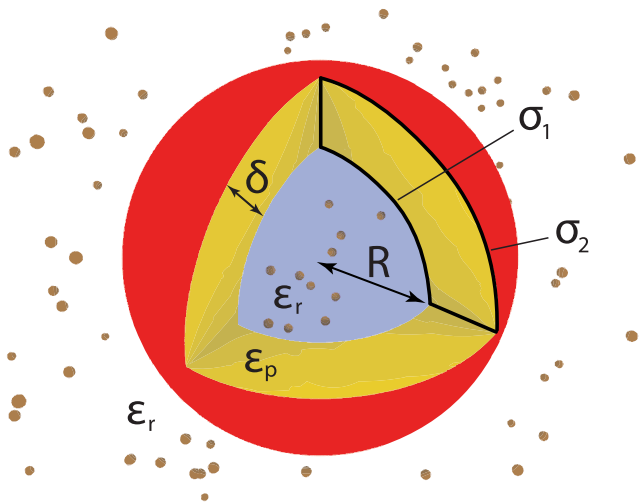,width=8.4cm}
}
\caption{An illustration of the electrostatic model of the viral capsid. The figure represents the 
cross-section of the assembled empty viral capsid with parameters of the 
model denoted. Salt ions are represented by small spheres. The interior of the 
capsid contains water and salt ions, but the salt ions are not present in the 
viral capsid shell.}
\label{fig:figempty4}
\end{figure}

In the electrostatic model from the previous section we have assumed 
that the capsid is infinitely thin and uniformly charged. The proteins of viral capsids 
carry a distribution of charge, both of positive and negative sign \cite{Karlin}, the negative charges 
being typically localized on the outside and the positive charges on the inside of the capsid (this trend is 
significantly more pronounced in DNA viruses \cite{Karlin}). The presented model 
could thus be viewed as a smeared lowest order (monopole) in multipole expansion of the fields generated 
by the protein charge distribution. In order to investigate whether the $R^2$ behavior of the electrostatic free 
energy of the capsid is a consequence of the simplicity of model I, in this section we adopt a more complex 
model of the capsid, as illustrated in Fig. \ref{fig:figempty4}. The capsid is approximated by the shell of finite thickness, 
$\delta$ whose dielectric permittivity is $\epsilon_p$. The shell is permeable to water and salt 
ions from the outside, but no ions are allowed in the capsid material. The inner and outer surfaces of the 
shell carry the surface charge densities 
of $\sigma_1$ and $\sigma_2$, respectively. The thickness of the dielectric layer $\delta$ in our model 
should not be identified with the capsid thickness, although they are clearly related. The typical capsid 
thicknesses are of the order of 2 nm \cite{Baker}, and this should be the upper bound for the $\delta$.

%%%%%%%%%%%%%%%%%%%%%
%%%%%%%%%%%%%%%%%%%%%%
It is again instructive to obtain some analytical limits for the energetics of viral capsids and 
to this end we solve the problem in the Debye-H\"{u}ckel approximation. The analytic results for 
the electrostatic potential are somewhat cumbersome and are thus summarized in the Appendix. The capsid free 
energy is in the DH approximation given by
\begin{equation}
F_{DH} = \frac{Q \Phi}{2} = 2 \pi \left [ \sigma_1 R^2 \Phi(R) + \sigma_2 (R+\delta)^2 \Phi(R+\delta) \right ].
\end{equation}
The number of parameters in this model is large, but since we are interested in the assembly of viral 
capsids, we seek for the expression for the free energy in the limit when $\kappa_{DH} R \gg 1$, 
$\delta \ll R$, and $\epsilon_r > \epsilon_p$ ($\epsilon_p$ is of the order of 5 for 
uncharged portions of the protein assemblies, depending also on the type of protein \cite{Pitera,Archontis}). 
In this regime, the free energy simplifies to 
\begin{equation}
\lim_{vir} F_{DH} = 2 \pi \frac{\epsilon_p (\sigma_1 + \sigma_2)^2 + \epsilon_r (\sigma_1^2 + \sigma_2^2) \kappa_{DH} \delta}
{\epsilon_0 \epsilon_r \kappa_{DH} (2 \epsilon_p + \epsilon_r \kappa_{DH} \delta )} R^2.
\label{eq:limdhmod2}
\end{equation}
Note that when $\delta=0$ the above equation reduces to Eq. (\ref{eq:limitlarge}) with $\sigma=\sigma_1+\sigma_2$. At 
physiological salt concentrations, $\kappa_{DH} \sim 1$ nm$^{-1}$, so that $\kappa_{DH} \delta \sim 1$, assuming 
that $\delta$ is not much smaller than 1 nm. Assuming that $\epsilon_r \gg 2 \epsilon_p$, a simpler expression for 
free energy of viral capsids is obtained: 
\begin{equation}
\lim_{\kappa_{DH} \delta \sim 1, \; \epsilon_r \gg \epsilon_p, \; \delta \ll R} 
F_{DH} = \frac{2 \pi (\sigma_1^2 + \sigma_2^2) R^2}{\epsilon_0 \epsilon_r \kappa_{DH}}.
\label{eq:DHdiele}
\end{equation}
Note that the above equation reduces to Eq. (\ref{eq:limitlarge}) when $\sigma_1 = \sigma_2 = \sigma / 2$, 
but for any other combination of $\sigma_1$ and $\sigma_2$ such that $\sigma_1 + \sigma_2 = \sigma$, the 
free energy is larger. Interestingly enough, the limiting form of the DH approximation for the free 
energy in Eq. (\ref{eq:DHdiele}) does not depend on $\epsilon_p$ as long as $\epsilon_p \ll \epsilon_r$ 
\cite{primjedba2}. Additionally, in this limit, the free energy does not depend on signs of 
$\sigma_1$ and $\sigma_2$.

\begin{figure}[ht]
\centerline{
\epsfig {file=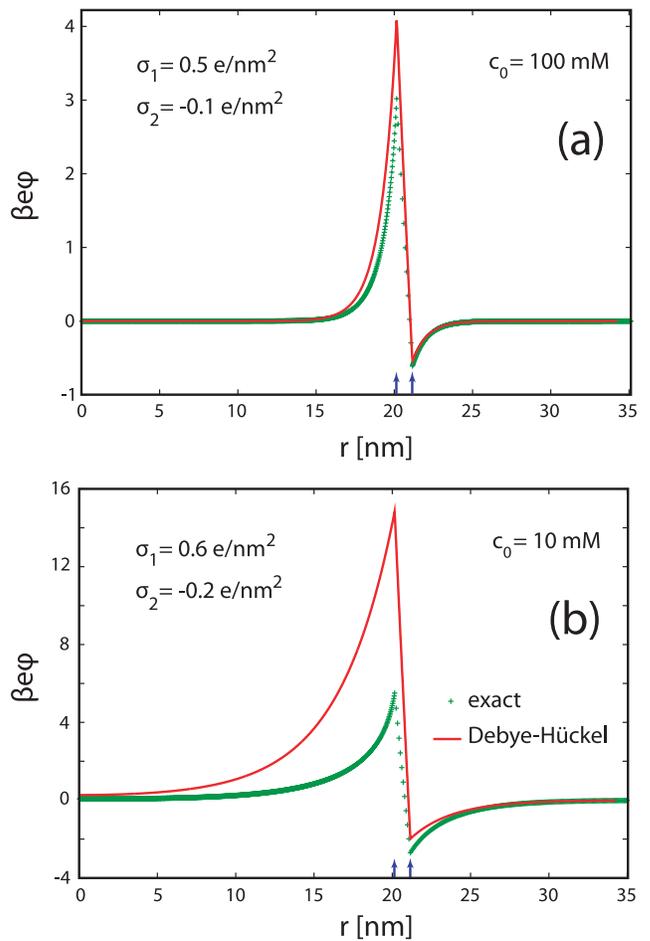,width=8.4cm}
}
\caption{Electrostatic potentials obtained numerically (symbols) and in the Debye-H\"{u}ckel approximation (lines). 
The arrows denote positions of the inner and outer side of the capsid. The parameters of the 
calculation are $\epsilon_p =5$, $\delta=1$ nm, and $R=20.15$ nm.
Panel (a): $\sigma_1=0.5$ $e$/nm$^2$, $\sigma_2=-0.1$ $e$/nm$^2$, $c_0=100$ mM. Panel (b): 
$\sigma_1=0.6$ $e$/nm$^2$, $\sigma_2=-0.2$ $e$/nm$^2$, $c_0=10$ mM}
\label{fig:figempty5}
\end{figure}

\begin{figure}[ht]
\centerline{
\epsfig {file=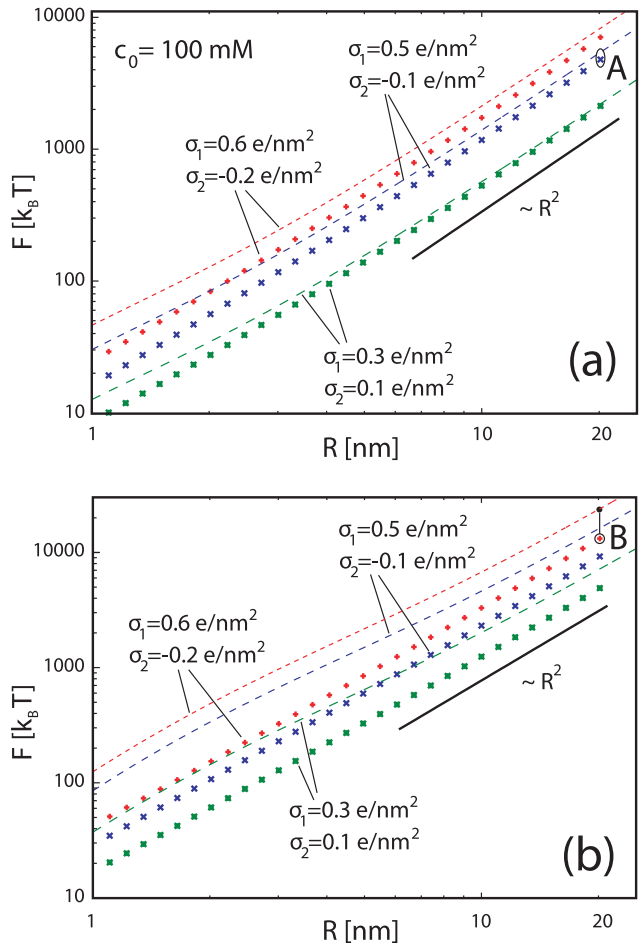,width=8.4cm}
}
\caption{Electrostatic free energies of the system (capsid + salt) obtained numerically (symbols) and in the Debye-H\"{u}ckel 
approximation (dashed lines) for three different combinations of the inner and outer surface charge densities, $\sigma_1$ and 
$\sigma_2$ as denoted in the figure. The parameters of the calculation are $\epsilon_p =5$ and $\delta=1$ nm. 
Panel (a): $c_0=100$ mM. Panel (b): $c_0=10$ mM. Full thick lines represent functions that are proportional to $R^2$. The 
electrostatic potentials corresponding to points denoted by A and B are presented in panels (a) and (b) of 
Fig. \ref{fig:figempty5}, respectively.}
\label{fig:figempty6}
\end{figure}

In any case, in the regime typical for viruses, we obtain the $R^2$ behavior of the free energy, as in model I. 
To check the validity of the DH approximation for this model, and whether the $R^2$ dependence of the free 
energy is preserved in the exact solution of the problem, we have again numerically solved the full nonlinear 
Poisson-Boltzmann equation. The results of these studies are displayed in 
Fig. \ref{fig:figempty5}, which represents the comparison of the DH and numerically exact solutions for the 
electrostatic potential, and in Fig. \ref{fig:figempty6} which displays the electrostatic energies of 
viral capsids.
%%%%%%%%%%%%%%%%%%%%%
%%%%%%%%%%%%%%%%%%%%%%

%%%%%%%%%%%%%%%%%%%%%
%%%%%%%%%%%%%%%%%%%%%%

It is obvious that the DH approximation may be entirely inapplicable to obtain reliable estimates for viral 
energetics, especially in the low to moderate salt regime. This is most easily seen by comparing the 
potentials obtained numerically and in the DH approximation in the low salt regime displayed in 
Fig. \ref{fig:figempty5}b) (note the scale of the potential). Note that the $R^2$ dependence of the free energy, 
predicted by the DH approximation, is confirmed by the 
numerical results in the limit appropriate for viral capsids. Intriguingly, inspection of the 
data obtained by numerically solving the problem reveals that the proportionality of free energy 
to square of capsid radius holds much better than predicted by the DH results, especially in the low-salt regime, and 
even in regime where $\kappa_{DH} R \sim 2$, at least for the parameters chosen in the presented results. This is 
similar to what was observed in model I. Note that the magnitudes of free energy obtained in model II 
for $\sigma_1 + \sigma_2=0.4$ $e$/nm$^2$ (2000 to 7000 $k_B T$ at $R=20$ nm, $c_0=100$ mM, 
depending on $\sigma_1$ and $\sigma_2$) are comparable but always larger than those obtained in 
model I for $\sigma = 0.4$ $e$/nm$^2$ ($\approx$ 1800 $k_B T$ at $R=20$ nm, $c_0=100$ mM), in 
agreement with Eq. (\ref{eq:DHdiele}) and the discussion following it.

\section{Application of the results to viruses}
\label{sec:application}

\subsection{$R^2$ dependence of free energy and preferred mean curvature of the capsid}
In the regime of radii typical for the viral capsids 
($R \sim$ 10 nm), $R \gg \lambda_{DH}$ limit is clearly appropriate for the physiological salt concentrations 
($c_0 \sim$ 100 mM) and thus $F \propto R^2$. The total free energy, i.e. the sum of 
free energies in Eqs. (\ref{eq:totalfree}) and (\ref{eq:adhesimp}) is also proportional to $R^2$, which 
means that the free energy {\em per protein} is a constant. This is confirmed by both of the 
models we considered. Since the effect related 
to entropic differences between the (free) proteins in the solution and those assembled 
in capsids are expected to be secondary for (large) protein capsids \cite{anomalousclust}, 
the electrostatic interactions in combination with the simple 
expression for the adhesion free energy [Eq. (\ref{eq:adhesimp})] cannot explain the occurrence 
of the preferred capsid radii. This is simply because irrespectively of the capsid size, i.e. 
its ${\cal T}$ number [see Eq. (\ref{eq:protnumber})], the free energy per protein remains 
constant. Thus, there is no particular capsid radius that is preferred energywise.
This is one of the main results of our study. 

A preferred mean curvature of the capsid ($H_0$) which 
could stem from the curvature of the protein-protein contacts would 
enforce a particular capsid radius - this would mean that the (non-electrostatic) 
''adhesion'' energy in Eq. (\ref{eq:adhesimp}) could be reformulated as 
\begin{equation}
F_{adhesion} = - \frac{N m}{2} \left[ \gamma + \omega \left(H_0 - \frac{1}{R} \right )^2 \right], 
\label{eq:adheadvanced}
\end{equation}
at least when $|H_0 - R^{-1}| \ll 1$. Similar views have been recently proposed 
\cite{bruinthermo,bruinspont}, but 
the original idea goes back to Caspar and Klug \cite{casparklug}. 
Our results clearly support the idea of the preferred curvature in viral capsids, a subject that 
has recently received much attention \cite{Siber1,bruinthermo,bruinspont,Nelson1}. 
If the attractive interaction has two or more minima in the spontaneous curvature, arising 
from several stable protein-protein conformations, one can also expect formation of 
several differently sized capsids \cite{zlotnickCCMV}. It should also be kept in mind that 
allowed radii of the capsid are in fact discrete, since the sphere can be 
triangulated by protein subunits only for certain total number of proteins [see 
Eq. (\ref{eq:protnumber})]. The allowed capsid 
radii are thus $R =  \sqrt{15 {\cal T}A_0/ \pi} $. 

Several articles 
have recently questioned the problem of pressure that acts within the 
self-assembled RNA virus \cite{bruin1,bruinpress}. In this respect loosely packed RNA 
viruses are expected to be very different from the double stranded DNA viruses whose 
capsids are known to withstand very high internal pressures \cite{Evilevitch,Purohit,Siber1}. 
It is important to 
note here that the fact that the empty capsids precisely assemble at radii given by 
$R = H_0^{-1}$, which is the same as the radius of the capsids filled with the viral 
genetic material (polyelectrolyte), means that the pressure that acts on the capsid arises 
solely from the polyelectrolyte self interaction and its interaction with the capsid \cite{bruinpress}. 
In other words, the repulsive electrostatic self-interactions in empty capsids can expected to be 
exactly counteracted by the 
angle-dependent adhesive interactions in Eq. (\ref{eq:adheadvanced}) - the first 
derivative of the sum of these interactions with respect to capsid radius (i.e. pressure) is zero.

\subsection{Influence of electrostatic interactions the strength of protein-protein contacts}

The capsid total free energy is negative only for sufficiently low 
values of the capsid charge density, i.e. when
\begin{equation}
\sigma^2 < \frac{2 m \gamma A_0 \epsilon_0 \epsilon_r}{\lambda_{DH}},
\end{equation}
where $A_0$ is the mean area of the protein subunit in the capsid [this estimate is based on 
the model I; see Eq. (\ref{eq:nosaltfree})]. This shows that for sufficiently large $\lambda_{DH}$ the 
repulsive electrostatic interaction dominates the energetics of the assembly. In 
solutions of low salt concentrations the capsid assembly should be thus inhibited. 
For sufficiently small values of adhesion constant $\gamma$, our calculations 
predict that there is a critically low salt concentration, $c_0^{cr}$ at which the assembly does not take 
place. Assuming that the critical salt concentration is high enough so that the limit 
$R / \lambda_{DH} (c_0^{cr}) \gg 1$ is still satisfied, one obtains that 
$c_0^{cr} = (2 \pi^2 \sigma^4 R^4 k_{B} T) / (\epsilon_0 \epsilon_r N^2 m^2 \gamma^2 e^2)$.

Although the presented electrostatic models cannot explain the preferred radii of the empty viral capsid, 
they can give some clue with regard to the recently observed strengthening of the protein-protein 
contacts in empty Hepatitis B capsids in high salt concentrations \cite{zlotnickhepb}. 
Namely, at high salt, the repulsive protein-protein electrostatic interactions are 
weakened as is clearly demonstrated in Fig. \ref{fig:figempty2}. Ceres and Zlotnick 
observed that the contact energy [$\gamma$, see Eq.(\ref{eq:adhesimp})] 
between the capsid proteins of Hepatitis B increases 
from -5.3 $k_B T$ at $c_0$=150 mM to -6.9 $k_B T$ at $c_0$=700 mM \cite{zlotnickhepb}. 
The total increase in energy of the capsid is thus $240 (-6.9 + 5.3)$ $k_B T$ = 384 $k_B T$, since 
the capsid is made of 120 copies of a tetravalent ($m=4$) protein dimer Cp149$_2$ \cite{zlotnickhepb} 
[see Eq. (\ref{eq:adhesimp})]. Our calculations based on model I predict that for $\sigma$=0.4 $e$/nm$^2$, 
the capsid free energies at $R$=15.6 nm (which is the mean radius of Hepatitis B capsid 
\cite{Wynne}), calculated from Eq. (\ref{eq:fefunc}) are 850 $k_B T$ 
and 415 $k_B T$ for $c_0$ = 150 mM and 700 mM, respectively. The calculated increase 
in the binding energy of a capsid is thus 435 $k_B T$ in a surprisingly good agreement with 
experimental results. Similar results are obtained in model II using 
e.g. $\sigma_1$=0.26 $e$/nm$^2$, $\sigma_2$=-0.13 $e$/nm$^2$, $\delta \approx 1$ nm, and 
$\epsilon_p = 5$ [compare Eqs. (\ref{eq:DHdiele}) and (\ref{eq:limitlarge})]. 
Obviously, quality of the agreement 
crucially depends on the value of the surface charge density adopted, since the free energy scales with the 
square of $\sigma$ [see Eq.(\ref{eq:freeDHtot}) and Fig. \ref{fig:figempty3}b)]. 
Assuming that $\sigma$=0.7 $e$/nm$^2$, as has 
been estimated in Ref. \onlinecite{Kegel} (see also the data estimated in 
Ref. \onlinecite{MuthuPNAS}), the increase 
in the binding energy is about 1220 $k_B T$, more than three times larger from the experimental 
estimate. Nevertheless, the order of magnitude agreement 
clearly suggest that the observed increase in the ''contact energy'' between the protein 
subunits with the increase of salt concentration should be ascribed to screening of the 
repulsive electrostatic protein-protein interactions rather than to some salt-induced 
conformational change in the protein structure as has been assumed in 
Ref. \onlinecite{zlotnickhepb}. The same conclusion has been found Ref. \onlinecite{Kegel}.

\subsection{Electrostatics and the inside/outside asymmetry of the viral capsids}

\begin{figure}[ht]
\centerline{
\epsfig {file=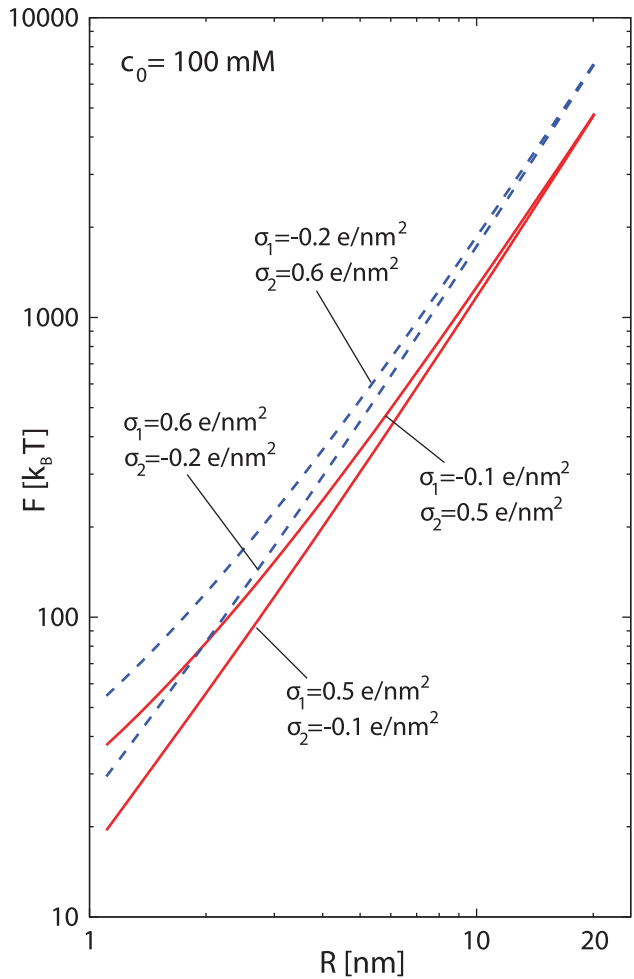,width=8.4cm}
}
\caption{Dependence of capsid free energies when the inner and outer charge densities 
interchange. Full lines denote the free energies when $\sigma_1 = 0.5$ $e$/nm$^2$ and 
$\sigma_2 = -0.1$ $e$/nm$^2$ ($\sigma_1 = -0.1$ $e$/nm$^2$ and 
$\sigma_2 = 0.5$ $e$/nm$^2$). Dashed lines denote the free energies when 
$\sigma_1 = 0.6$ $e$/nm$^2$ and $\sigma_2 = -0.2$ $e$/nm$^2$ 
($\sigma_1 = -0.2$ $e$/nm$^2$ and $\sigma_2 = 0.6$ $e$/nm$^2$). In these calculations, 
$\epsilon_p=5$, $\delta=1$ nm, and $c_0=100$ mM.}
\label{fig:figempty7}
\end{figure}

Although the asymptotic behavior of energies in the DH approximation [Eq. (\ref{eq:DHdiele}] 
does not show the difference in cases when the inner charge density becomes the outer and 
vice versa, numerically exact results do show a slight difference in 
the regime appropriate for viral capsids, as illustrated in Fig. \ref{fig:figempty7}.
%%%%%%%%%%%%%%%%%%%%%
%%%%%%%%%%%%%%%%%%%%%%
The free energies are smaller in the case when larger charge density is on the inner side of the 
capsid. Note, however, that the difference in free energies in the two cases is 
less than about 100 $k_B T$ per capsid, at least for the parameters chosen for the 
calculations displayed in Fig. \ref{fig:figempty7}. This translates in about $k_B T$ per 
protein subunit, which is of the order of 10 \% of the total energy of 
protein-protein contacts. Thus, the inside/outside (top/bottom) asymmetry of the protein charge 
may have some influence on the precision of the assembly so that the proper sides of proteins 
face the capsid interior. Again, preferred curvature and 
steric constraints could be more important reasons for the precision of viral assembly.

\section{Limitations of the models}
\label{sec:seclimit}

Our models of an empty viral capsid are approximate and limited in several aspects. First, we have assumed 
that the surface charges are distributed uniformly on infinitely thin spheres. This is of course 
an oversimplification of the true distribution of protein charges in the capsid. Model 
II does account for the essential characteristics of the capsid charge distribution, but it 
still does not incorporate the posible polar and azimuthal inhomogeneities of capsid charge. In that 
respect, it is of interest to note that the $R^2$ behavior of the free energy obtained in model I 
for $\kappa_{DH} R \gg 1$ is obtained, at least on the level of the DH approximation, even 
when one accounts for the charge inhomogeneities on the surface of the sphere that have the icosahedral 
symmetry of the capsid \cite{Marzec}. Thus, it doesn't seem likely that the spatial distribution of 
protein charge could favor the preferred angle of protein-protein contacts. 

A second approximation in our models is the assumption of perfect sphericity of the viral capsid. 
The elasticity model of viral capsids by Lidmar {\sl et al} \cite{Nelson1} suggests that larger 
viruses are more facetted and that their asphericity is larger. This is indeed a general 
trend that is observed in the experimental studied of viruses \cite{Baker}. However, 
the strict predictions of the theory are based on its continuum limit (i.e. large ${\cal T}$ 
numbers) which may not hold for small viruses. Nevertheless, numerical 
experiments with model capsid of smaller ${\cal T}$-numbers \cite{Siber1,Siber2}, performed 
as detailed in Refs. \onlinecite{Nelson1,Siber1}, do show that the asphericity in small viruses 
is again governed by the interplay of bending and stretching energies of the viral capsid and by 
the capsid radius. Elastic properties of such small capsids that are very much different from 
the corresponding properties of large viral capsids would be 
needed to reproduce a significant asphericity in small viruses, knowing the asphericity of 
larger viruses where the continuum limit of the theory is expected to work fine. Thus, the 
assumption of small asphericity in capsids of small viruses is corroborated by the 
discrete version of the models elaborated in Refs. \onlinecite{Nelson1,Siber1}. This, however 
does not account for the corrugation of the capsid surface that is observed experimentally, 
especially in some viruses (e.g. bacteriophage P4, see Ref. \onlinecite{Baker}). This effect 
is related to the spatial distribution of the protein mass and charge, and, 
as elaborated in the previous paragraph, is neglected in our model.

\section{Summary}

In summary, we find that electrostatic interactions make important (repulsive) 
contribution to the energetics empty capsid binding of several hundreds of 
$k_B T$, depending strongly on the salt concentration. In highly salty solutions, 
the electrostatic interactions are efficiently screened resulting in a
larger binding energy of the capsids, in accordance with what was experimentally 
found for Hepatitis B virus \cite{zlotnickhepb}. The results of our exact numerical 
studies compare favorably with limiting expressions derived previously in the 
Debye-H\"{u}ckel approximation and valid for sufficiently large salt concentrations \cite{Kegel}. 
The regimes in which these approximations severely fail have been identified. 
Since the electrostatic energies of viral capsids scale with the second power of 
capsid radius, in the regime of radii typical for viruses, we find that 
simple, curvature independent expressions for the adhesive, 
attractive interaction, such 
as Eq. (\ref{eq:adhesimp}), cannot 
explain the monodispersity of self-assembled empty viral capsids and that certain 
angle-dependent interaction acting between the neighboring protein subunits is 
needed in that respect. Our study also explains that if the 
empty viral structure that is formed spontaneously has the same symmetry as fully 
infectious virus (containing the genetic material/polyelectrolyte) then the 
pressure that acts on the viral capsid arises from the polyelectrolyte 
self-interaction, and the attractive interaction between the polyelectrolyte and 
the viral capsid, {\sl i.e.} the capsid electrostatic self-repulsion is exactly counteracted by 
the curvature dependent adhesion energy.

\section{Acknowledgments}

This work has been supported by the Agency for Research and Development of Slovenia 
under grant P1-0055(C), the Ministry of Science, Education, and Sports of Republic of Croatia through Project No. 035-0352828-2837, 
and by the National Foundation for Science, Higher Education, and 
Technological Development of the Republic of Croatia through Project No. 02.03./25. 

\appendix*
\section{Details of the solution of model II in the Debye-H\"{u}ckel approximation}

The potential in the interior of the capsid can in the DH approximation be written as 
\begin{equation}
\Phi(r) = A \frac{\sinh ({\kappa_{DH}} r)}{r}, \; r<R.
\end{equation}
In the capsid material, the potential is 
\begin{equation}
\Phi(r) = \frac{C}{r} + D, \; R<r<R+\delta,
\end{equation}
and outside the capsid
\begin{equation}
\Phi(r) = B \frac{\exp(-\kappa_{DH} r)}{r}, r>R + \delta.
\end{equation}
Two equations for four unknown coefficients, $A$,$B$,$C$ and $D$ are obtained from the requirement 
of the continuity of potential at $r=R$ and $r=R + \delta$. Additional two equations are 
obtained by relating the discontinuity in the electric displacement field to the surface charge 
density,
\begin{eqnarray}
\epsilon_0 \epsilon_r \left. \frac{\partial \Phi}{\partial r} 
\right |_{r-R=0^-} - \epsilon_0\epsilon_p \left.\frac{\partial \Phi}{\partial r} \right |_{r-R=0^+} &=& \sigma_1 \nonumber \\
\epsilon_0 \epsilon_p \left. \frac{\partial \Phi}{\partial r} 
\right |_{r-R-\delta=0^-} - \epsilon_0\epsilon_r \left.\frac{\partial \Phi}{\partial r} \right |_{r-R-\delta=0^+} &=& \sigma_2 . 
\end{eqnarray}
By solving the thus obtained equations, one obtains fairly complicated expressions for the unknown 
coefficients, which we simplify somewhat by introducing substitutions
\begin{equation}
A = {\cal A} / {\Delta}, \;
B = {\cal B} / {\Delta}, \;
C = {\cal C} / {\Delta}, \;
D = {\cal D} / {\Delta},
\end{equation}
where
\begin{eqnarray}
\Delta &=& \epsilon_0 \epsilon_r \left \{ (\epsilon_p - \epsilon_r) \kappa_{DH} \delta^2 + \epsilon_p \kappa_{DH} R^2 \right. \nonumber \\
&+& \left[ \epsilon_p (1+ 2 \kappa_{DH}R) - \epsilon_r (1 + \kappa_{DH} R) \right] \delta  \nonumber \\
&+&  \kappa_{DH} R \left[ \epsilon_r \delta^2 \kappa_{DH} + \epsilon_p R \right. \nonumber \\
&+& \left . \left . \epsilon_r \left( 1 + \kappa_{DH}R \right ) \delta
\right] \coth (\kappa_{DH} R)
\right \},
\end{eqnarray}
and 
\begin{eqnarray}
{\cal A} &=& \left \{ \epsilon_r  \sigma_1  R^2 \kappa_{DH} \delta^2 + \epsilon_p  \left[\sigma_1 R^2 + \sigma_2 (R + \delta)^2 
\right] R \right . \nonumber \\ 
&+& \left. \epsilon_r  \sigma_1  (1 + \kappa_{DH} R) \delta R^2 \right \} \rm{csch} (\kappa_{DH} R),
\end{eqnarray}

\begin{eqnarray}
{\cal B} &=& \left \{ [\epsilon_p (\sigma_1 R^2 + \sigma_2 (R + \delta)^2) - \epsilon_r \sigma_2 (R + \delta)^2 ] \delta \right. \nonumber \\
&+& \epsilon_p [\sigma_1 R^2 + \sigma_2 (R + \delta)^2] R \nonumber \\
&+& \left. \epsilon_r \sigma_2 (R + \delta)^2 \delta \kappa_{DH} R \coth (\kappa_{DH} R)
\right \} \nonumber \\
&\times& \exp \left[ \kappa_{DH} (R + \delta) \right],
\end{eqnarray}

\begin{eqnarray}
{\cal C} &=& \epsilon_r \left \{ \sigma_1  \delta^2 \kappa_{DH} R^2 +  [\sigma_1 R^2 + \sigma_2 (R+\delta)^2] R \right. \nonumber \\
&+& \sigma_1 R^2 (1+2 \kappa_{DH} R)  \delta \nonumber \\ 
&-& \left.  \sigma_2 (R + \delta)^2 \kappa_{DH} \coth (\kappa_{DH} R)
\right \},
\end{eqnarray}

\begin{eqnarray}
{\cal D} &=& \epsilon_p [\sigma_1 R^2 + \sigma_2 (R+\delta)^2] \nonumber \\
&-& \epsilon_r \left [ \sigma_1 R^2 + \sigma_2 (R+\delta)^2 + \sigma_1 R^2 \kappa_{DH} (R + \delta) \right] \nonumber \\
&+& \epsilon_r \kappa_{DH} \sigma_2 (R + \delta)^2 R \coth (\kappa_{DH} R).
\end{eqnarray}

When $\delta = 0$, the potential reduces to that of a single shell of charge with the effective surface 
charge density of $\sigma = \sigma_1 + \sigma_2$ [in the DH approximation, see Eqs. (\ref{eq:DHpotential0}) and 
(\ref{eq:DHpotential})]. When $\kappa_{DH} R \gg 1$, $\delta \ll R$, and $\epsilon_r > \epsilon_p$, which is a regime 
of interest for viral capsids (denoted by subscript $vir$ in the equations below), the following 
limiting forms for $\Delta$, ${\cal A}$, and ${\cal B}$ 
coefficients apply:
\begin{equation}
\lim_{vir} \Delta = \epsilon_0 \epsilon_r \left[ \kappa_{DH} R^2 ( \epsilon_r \kappa_{DH} \delta + 2 \epsilon_p)  \right],
\end{equation}
\begin{equation}
\lim_{vir} {\cal A} = R^3 \left[ (\sigma_1 + \sigma_2) \epsilon_p + \sigma_1 \epsilon_r \kappa_{DH} \delta \right] \rm{csch}(\kappa_{DH} R),
\end{equation}
\begin{equation}
\lim_{vir} {\cal B} = R^3 \left[ (\sigma_1 + \sigma_2) \epsilon_p + \sigma_2 \epsilon_r \kappa_{DH} \delta \right] \exp(\kappa_{DH} R).
\end{equation}
These three limiting forms are sufficient to derive Eq. (\ref{eq:limdhmod2}).


\begin{thebibliography}{99}
\bibitem{primjedba} For calculations with $R>12$ nm, about 1200 points were used, 
which was important esspecially for high salt concentration. In that case, strong 
variation of the potential takes place close to the capsid and very small steps in radial coordinate are 
needed to accurately account for this effect.
\bibitem{primjedba2} The value of the dielectric constant of proteins has been a subject of controversy, see e.g. Refs. 
\onlinecite{Pitera,Archontis}. For our purposes, we approximate viral proteins essentially as disks of finite thickness with 
charges concentrated on the disk bases. Therefore, the dielectric constant that is of interest to us corresponds to 
the permittivity of the uncharged interior of the protein. It follows from Eq. (\ref{eq:limdhmod2}) that 
the value of $\epsilon_p$ cannot change the $R^2$ behavior of the electrostatic free energy.
\end{thebibliography}
\end{document}